\documentclass[pre,twocolumn]{revtex4-1}
\usepackage{amsmath}
\usepackage{graphicx}
\usepackage{hyperref}

\usepackage{color}
\newcommand{\red}[1]{{#1}}

\begin{document}

\title{Influence of Multiple Sequence Alignment Depth on Potts Statistical Models of Protein Covariation}

\begin{abstract}
Potts statistical models have become a popular and promising way to analyze mutational covariation in protein Multiple Sequence Alignments (MSAs) in order to understand protein structure, function and fitness. But the statistical limitations of these models, which can have millions of parameters and are fit to MSAs of only thousands or hundreds of effective sequences using a procedure known as inverse Ising inference, are incompletely understood. In this work we predict how model quality degrades as a function of the number of sequences $N$, sequence length $L$, amino-acid alphabet size $q$, and the degree of conservation of the MSA, in different applications of the Potts models: In ``fitness" predictions of individual protein sequences, in predictions of the effects of single-point mutations, in ``double mutant cycle" predictions of epistasis, and in 3-d contact prediction in protein structure. We show how as MSA depth $N$ decreases an ``overfitting" effect occurs such that sequences in the training MSA have overestimated fitness, and we predict the magnitude of this effect and discuss how regularization can help correct for it, use a regularization procedure motivated by statistical analysis of the effects of finite sampling. We find that as $N$ decreases the quality of point-mutation effect predictions degrade least, fitness and epistasis predictions degrade more rapidly, and contact predictions are most affected. However, overfitting becomes negligible for MSA depths of more than a few thousand effective sequences, as often used in practice, and regularization becomes less necessary. We discuss the implications of these results for users of Potts covariation analysis.
\end{abstract}

\author{Allan Haldane}
\email[Corresponding author: ]{allan.haldane@temple.edu}
\affiliation{Center for Biophysics and Computational Biology, Department of Physics, and Institute for Computational Molecular Science, Temple University, Philadelphia, Pennsylvania 19122}
\author{Ronald M. Levy}
\affiliation{Center for Biophysics and Computational Biology, Department of Chemistry, and Institute for Computational Molecular Science, Temple University, Philadelphia, Pennsylvania 19122}
\maketitle

\section{Introduction}

Potts models are statistical models with a rich history of study in condensed matter physics, and which more recently have found important applications in protein physics. Potts models can be parameterized from a Multiple Sequence Alignment (MSA) of a protein family to model the sequence likelihoods and pairwise amino-acid correlations observed in the MSA\cite{2IQu,t2aN,W5j0,9t9z}, with numerous uses relating protein structure, function and fitness. These models also have a rich interpretation in the language of biological and statistical physics through their relation to lattice models of protein folding \cite{NBbG,qEP2}, and can model other biophysical systems and datasets involving large numbers of correlated and interconnected components, such as networks of neurons\cite{rSDu}.

Potts models parametrized on MSAs of a protein family, using the procedure known as inverse Ising inference, have been shown to predict experimental measurements of proteins. Some predictions use the Potts ``statistical energy" of individual sequences, computed by adding up Potts ``coupling" parameters for all position-pairs for an individual sequence as outlined below, which reflects the likelihood of the sequence appearing in the MSA. Statistical energies have been used to predict sequence-dependent fitnesses \cite{bgwj}, enzymatic rates \cite{W5Kp}, melting temperature\cite{qEP2,WG7P}, and mutation effects \cite{7TfN}. Potts models can also be used to predict contacts in protein structure, as the coupling parameters of the model can indicate which position-pairs have the strongest ``direct" statistical dependencies, which are found to be good predictors of contacts \cite{toV5}. This contact information has been found to be enough to perform accurate ab-initio protein structure prediction from sequence variation data alone \cite{veQd,NKK1}.

Despite these advances, a more complete picture of the nature of the statistical errors inherent in Potts models of sequence co-variation and observables derived from them is lacking. The purpose of the present analysis is to further explore how MSA depth (number of sequences), MSA sequence length, amino-acid alphabet size, and other quantities determine model quality. In particular, how many sequences are necessary to give accurate contact predictions or fitness predictions? How does the model behave if too few sequences are provided? Furthermore, a Potts model for a typical protein family with sequence length $L=200$ and amino-acid alphabet of $q=21$ letters (20 amino-acids plus gap) has almost $10^7$ parameters, yet is fit to a relatively small number of sequences, often 100 to 10000 effective sequences, out of a sequence space of $\sim 21^{200}$ possible sequences. These large differences in scale raise the question of overfitting.

The effect of the MSA depth on model inference has been previously examined in some situations. One of the most detailed treatments of Potts statistical error is \cite{ZHvj}, where it is argued that sampling noise caused by small MSA depth can lead a well-conditioned Ising problem to become ill-conditioned, meaning that model parameters become sensitive to small changes in the MSA. This study also performed numerical tests, using the Adaptive Cluster Expansion inference algorithm, of the effect of sampling noise on certain model properties, though not on contact prediction or fitness prediction. These authors suggest that $l_1$ regularization helps correct for sampling noise if the interaction network is sparse\cite{wa7p,ZHvj}. In another study using mean-field inference methods, inference was tested for varying MSA depths of 72, 296, 502, 1206, and 2717 effective sequences, finding, for example, that the top 24 contacts were predicted at an 80\% true-positive rate for MSA depth of 296, which increases to 70 contacts for a depth of 1206 for the RAS family \cite{Fz8l}. As the sequence diversity and depth of the MSA are simultaneously decreased, the power of the Potts model has been found to decrease, both for mutation-effect predictions using a pseudolikelihood inference method\cite{7TfN}, and contact prediction \cite{8aHM}. However, these results do not give a clear view of the statistical errors due to finite sampling alone because of the presence of various non-statistical forms of error or bias.

\red{It is useful to recall these other potential biases of covariation analysis in order to distinguish them from finite-sampling error. These biases vary from study to study. During MSA construction biases arise due to choices in the diversity cutoff of collected sequences, how to account for gap characters, and how to align the sequences \cite{Rve7}. It is then common to downweight similar sequences in the MSA to account for phylogenetic structure, which can have a significant effect on the estimated residue frequencies and lowers the ``effective" number of sequences of the MSA, dependent on the choice of similarity threshold. During inference biases can arise due to various approximations used to speed up the inference algorithm at the cost of accuracy, and also due to the choice of regularization strategy. Strong regularization has been shown to be essential when using the more approximate mean-field inference algorithm \cite{ez9b}, and regularization has been found to affect the ``flatness" of the inferred fitness landscape of Potts-like models inferred by linear regression\cite{xrrT}. There are also potentially biases due to model mis-specification due to the absence of higher-order coupling terms in the model, however there is evidence that Potts models accurately describe the complex mutational statistics of real protein families\cite{xIOL}.

Finite sampling error is a fundamental statistical limitation of all inverse Ising inference implementations. The goal of the present study is to clarify the limitations of Potts model inference due uniquely to finite-sampling effects. In support of this goal we use a model inference algorithm which avoids analytic approximations and which has been shown to accurately reproduce the sequence mutational statistics when used to generate new sequences \cite{8K1h,xIOL,Cmda}, and focus on three types of model predictions: Statistical energy predictions of individual sequences, mutation-effect predictions including predictions of ``double mutant cycle" tests of epistasis \cite{kgPd,Ado2}, and contact prediction.}

We derive the expected correlation coefficient $\rho(E, \hat{E})$ between the benchmark statistical energies of sequences and their estimate based on a Potts model fit to a finite-depth MSA as a function of the MSA parameters $N$, $L$, $q$, and the degree of conservation, using a simplified model. We also illustrate how overfitting occurs for small MSAs, which lowers (makes more favorable) the predicted statistical energy of the sequences in the training MSA relative to sequences which were not in the training MSA. This effect is relevant when comparing the statistical energy of different sequences, particularly for small MSAs, and we discuss whether this affects common calculations such as predictions of the fitness effects of mutations. While the quality of all types of Potts model predictions degrades as the MSA depth $N$ decreases, the predictions of point-mutation effects are the least affected and give high correlations to the reference values even for very small MSAS. 

We verify these results for the Potts model using \textit{in-silico} numerical tests. We use two protein families in our numerical tests, the protein kinase catalytic domain family, and the SH3 family. These families are of particular interest to us biologically, but here we use them as example systems with which to test and demonstrate the statistical properties of inverse Ising inference, particularly because of the wealth of sequence and structural information on them.

\section{Background and Methods}

\subsection{Potts Models}

Explanations of how Potts models are used in protein sequence covariation analysis have been presented in many previous studies \cite{lxcT,OatS,ZHvj}, and we summarize the relevant aspects here. A Potts model, in this context, is the maximum-entropy model for the probability $P(S)$ of sequences in a protein family, constrained to predict the pairwise (bivariate) amino-acid (residue) frequencies $f^{ij}_{\alpha \beta}$ of an MSA of that family, for residues $\alpha,\beta$ at pairs of positions $i,j$. \red{These bivariate marginals are computed from a given MSA by counting the number of times each residue-pair is present, or
\begin{equation}
f^{ij}_{\alpha\beta} = \frac{1}{N} \sum_{S \in \text{MSA}} \delta^{\alpha}_{S_i} \delta^{\beta}_{S_j}.
\end{equation}}

Given an MSA of sequence length $L$ and alphabet of $q$ letters, there are ${L \choose 2} q^2$ bivariate frequencies used as model constraints, although because the univariate frequencies $f^i_\alpha = \sum_\beta f^{ij}_{\alpha \beta} = \frac{1}{N} \sum_{S \in \text{MSA}} \delta^{\alpha}_{S_i}$ must be consistent across all pairs and sum to 1 the constraints are not independent, and can be reduced to ${L \choose 2} (q-1)^2$ bivariate plus $L(q-1)$ univariate independent constraints. Maximizing the entropy with these constraints leads to an exponential model in which the likelihood of the dataset MSA is $\mathcal{L}(\text{MSA}) = \prod_{S \in \text{MSA}} P(S)$, a product of sequences probabilities with distribution $P(S) = e^{-E(S)}/Z$, with a ``statistical energy" 
\begin{equation}
E(S) = -\sum_i h^i_{S_i} - \sum_{i<j} J^{ij}_{S_i S_j}
\label{eq:statistical_energy}
\end{equation}
which is a sum over position- $(i,j)$ and residue- $(s_i, s_j)$ specific ``coupling" parameters $J^{ij}_{S_i S_j}$ and ``field" parameters $h^i_\alpha$ to be determined from the data, with a ``partition function" $Z = \sum_S e^{-E(S)}$. The couplings $J^{ij}_{\alpha \beta}$ can be thought of as the statistical energy cost of having residues $\alpha,\beta$ at positions $i,j$ in a sequence. Given a parameterized model one can generate new sequences from the distribution $P(S)$, for instance using Monte-Carlo methods.

In equation \ref{eq:statistical_energy} we have simplified the notation by defining ${L \choose 2} q^2$ coupling parameters and $Lq$ fields, however because of the non-independence of the bivariate marginal constraints some of these are superfluous. One can apply ``gauge transformations" $(h^i_\alpha, J^{ij}_{\alpha \beta}) \rightarrow (h^i_\alpha + a^i + d^i_\alpha, J^{ij}_{\alpha \beta} + b^i + c^j - d^i_\alpha)$ for arbitrary constants $a^i, b^i, c^j, d^i_\alpha$ and this does not change the probabilities $P(S)$ and only results in a constant energy shift of all sequences. By imposing additional gauge constraints one finds the model can be fully specified using the same number of parameters $\theta^\text{Potts} = {L \choose 2}(q-1)^2 + L(q-1)$ as there are independent marginal constraints. A common choice of gauge constraint is to fix $J^{ij}_{q \beta} = -\sum_{\alpha \neq q} J^{ij}_{\alpha \beta}$ and $h^i_q = -\sum_{\alpha \neq q} h^i_\alpha$, called the ``zero-mean gauge" since the mean value of the couplings and fields is 0, and this is the gauge which minimizes the squared sum of the couplings, or Frobenius Norm, $\text{FB}(i,j) = \sum_{\alpha \beta} (J^{ij}_{\alpha\beta})^2$ whose use is described below. It is also possible to transform to a ``fieldless" gauge in which all of the fields $h^i_\alpha$ are set to 0, which is sometimes computationally convenient.

By fitting the bivariate frequencies this model captures the statistical dependencies between positions, whose strength is reflected in the correlations $C^{ij}_{\alpha\beta} = f^{ij}_{\alpha\beta} - f^i_\alpha f^j_\beta$. Importantly the model allows us to distinguish between ``direct" and ``indirect" statistical dependencies, which is not possible based on the $C^{ij}_{\alpha\beta}$ directly. The directly dependent pairs are defined by ``strong" (nonzero) couplings $J^{ij}_{\alpha \beta}$ in the Potts model, and networks of strong couplings $J^{ij}_{\alpha \beta}$ can cause indirect and higher-order statistical dependencies, even though the couplings are only pairwise. This is useful because position-pairs with strong direct couplings have been shown to best reflect 3D contacts in protein structure. For each position pair one can estimate the strength of the direct statistical dependence between a pair of positions by various ``direct interaction scores", for example with the Frobenius Norm in the zero-mean gauge. A common feature of these scores is that if the couplings $J^{ij}_{\alpha\beta} = 0$ in the zero-mean gauge, then there is no direct dependency, even if $C^{ij}_{\alpha\beta}$ is nonzero.

One can also compute the Potts statistical energy $E(S)$ for any sequence. The statistical energy reflects how likely a sequence is to appear in the MSA, which is expected to relate to evolutionary ``fitness". \red{While protein fitness is a function of many molecular phenotypes, it is sometimes hypothesized to be dominated by the requirement that the protein folds, in which case the Potts statistical energy of a sequence is expected to correlate well with its thermostability. Experimental measurements of the thermostability of some proteins have been found to correlate well with $E(S)$ \cite{WG7P,3LtB,W5Kp,wiT3,BRO2}.} A common application of the statistical energy score is to predict the fitness effect of a point-mutation to a sequence through the change in statistical energy $\Delta E$ it causes. A point-mutation causes a change in $L$ of the coupling values for that sequence, and the collective effect of the pairwise coupling terms appears to be crucial for correctly predicting sequence fitnesses \cite{xIOL}.

\subsection{The Independent Model}

We contrast the Potts model with the ``independent" model, the maximum-entropy model for $P(S)$ constrained to reproduce only the MSA's single-site residue frequencies $f_\alpha^i$. It takes the form $P(S) \propto e^{-E(S)}$ with a ``statistical energy" $E(S) = -\sum_i^L h_{s_i}^i$. Unlike the Potts model the independent model is separable and $P(S)$ can be written as a product over positions $P(S) \propto \prod_i e^{h^i_{s_i}}$, and maximum likelihood parameters given an MSA are $h_\alpha^i = \log f_\alpha^i$. Even though the independent model does not capture statistical dependencies between positions like the Potts model it is in the same exponential family and behaves similarly in many respects. There are $\theta^\text{Indep} = L(q-1)$ independent univariate marginal constraints and an equal number of free field parameters after the gauge is constrained, analogously to the Potts model.

\subsection{Correlation Energy Terms}

Here we introduce a new quantity which will be used below, which we will call the ``correlation energy" and is given by
\begin{equation}
X^{ij} = -\sum_{\alpha\beta} J^{ij}_{\alpha\beta} C^{ij}_{\alpha\beta}
\end{equation}
for each position-pair $i,j$. We also define the ``total correlation energy" as $X = \sum_{ij} X^{ij}$.

These terms have the following useful interpretation. If we compute the mean statistical energy of sequences in the input MSA, and then create a new ``shuffled" MSA by randomly shuffling each column of the MSA, thus breaking any correlations between columns, and compute the mean statistical energy of these shuffled sequences, then the total correlation energy is equal to the mean difference, or energy gap, between these two sets of sequences. In this way the total correlation energy can be interpreted as the average statistical energy gained due to mutational correlations.  Another way to view it is as the mean Potts statistical energy difference between sequences generated by the Potts model, and sequences generated by the independent model, as mathematically $\sum_{ij} X^{ij} = (-\sum_{ij\alpha\beta} J^{ij}_{\alpha\beta} f^{ij}_{\alpha\beta}) - (-\sum_{ij\alpha\beta} J^{ij}_{\alpha\beta} f^i_\alpha f^j_\beta) = \langle E(S) \rangle_\text{Potts}  -  \langle E(S) \rangle_\text{Indep}$, using a fieldless gauge. The pairwise terms $X^{ij}$ can similarly be interpreted as the statistical energy gained due to correlations between columns $i$ and $j$ only.

These correlation energy terms have two important properties. First, they are gauge-independent, or invariant under the gauge transformations described above, since the rows and columns of the correlation matrices $C^{ij}_{\alpha\beta}$, shaped as $q \times q$ for each pair $i,j$, sum to 0. Second, they can be used as a measure of the strength of direct interaction between columns $i$ and $j$: For uncoupled pairs where $J^{ij}_{\alpha\beta} = 0$ in the zero-mean gauge, the correlation energy $X^{ij}$ will be 0, as expected. This score can be compared to other direct interaction scores, such as the ``Direct Information" \cite{5Pz1} or Frobenius norm. The correlation energy terms are attractive because they are both gauge-independent and have a simple interpretation in terms of the statistical energy of the sequences. We find that they are less accurate when used for contact prediction, but suggest they may better reflect the magnitude of the effects of residue-pair interactions on the fitness of mutants. We make use of the correlation energy terms to track convergence of the inverse Ising procedure and for regularization.

\subsection{Inverse Ising Inference From an MSA}

In this study we parametrize the Potts model using a Monte-Carlo GPU-based method \cite{8K1h}. Given a dataset MSA we aim to maximize the scaled log likelihood $\ell = \frac{1}{N} \log \mathcal{L}(\text{MSA}) = \sum_{ij} J^{ij}_{\alpha\beta} \hat{f}^{ij}_{\alpha\beta} - \log(Z)$ where $\hat{f}^{ij}_{\alpha\beta}$ are the dataset bivariate frequencies. The gradient of this log likelihood is $\frac{\partial \ell}{\partial J^{ij}_{\alpha_\beta}} = \hat{f}^{ij}_{\alpha\beta}  - f^{ij}_{\alpha\beta} \equiv \Delta f$, so the likelihood is minimized when bivariate marginal discrepancy $\Delta f$ is  $0$. We use a quasi-Newton numerical method to find this minimum, and estimate the model bivariate frequencies $f^{ij}_{\alpha\beta}$ given trial couplings by generating large simulated MSAs by parallel Markov Chain Monte Carlo (MCMC) over the landscape $P(S)$, and then update the couplings based on the discrepancy with the dataset MSA bivariate frequencies. We have implemented this algorithm for GPUs \cite{8K1h}.

This method avoids analytic approximations, though it is limited by the need for the MCMC procedure to equilibrate and by sampling error in the simulated MSAs. To minimize this ``internal" sampling error we use simulated MSAS of 1048576 sequences. We measure equilibration of each round of MCMC sequence generation by making use of this large number of parallel MCMC replicas, where each replica evolves a single sequence in time. Equilibration of the replicas is achieved once the autocorrelation of the replica energies for half the number of steps, $\rho(\vec{E}(t), \vec{E}(t/2))$, is uncorrelated with p-value of 0.02 or more. There is a second form of equilibration, of the model parameter values themselves over the course of multiple rounds of MCMC sequence generation, which we measure through the stationarity or leveling off of the total correlation energy $X$ defined above. In some cases, for instance for very small unregularized MSA datasets, the inference procedure failed to equilibrate in a reasonable time, as we discuss in results.

Because it makes no analytic approximations, this method leads to a model which can be used to generate simulated MSAs which accurately reproduce the dataset bivariate marginals and correlations, and we have previously shown also reproduces the higher-order marginals (corresponding to probabilities of subsequences of more than two positions) \cite{xIOL}. This generative property of the MCMC inference algorithm is key to our results below, as we wish to generate MSAs of varied depths $N$ whose statistics match the original dataset statistics up to finite sampling limitations. Our GPU implementation allows us to efficiently generate large simulated MSAs given a parameterized Potts model, which we use to perform statistical tests on the quality of Potts model inference using sampled MSAs.

\subsection{Overfitting}

It is well known that statistical models may ``overfit" due to finite sampling effects when the number of samples in the dataset used to parametrize the model is small. Overfitting of the Potts model parameters is ultimately due to the statistical error caused by finite sampling in the bivariate frequencies $f^{ij}_{\alpha\beta}$ used as input to the inference procedure, which are computed from the MSA of $N$ sequences. Each bivariate marginal $\hat{f}^{ij}_{\alpha\beta}$ is estimated from a sample of size $N$, and its statistical error is reflected by the multinomial mean-squared-error $\sigma^2 = f^{ij}_{\alpha\beta}(1-f^{ij}_{\alpha\beta})/N$. Since the bivariate marginals are the input into the inverse Ising algorithm, this statistical error in the inputs leads to error in the inferred parameters. We note that overfitting is not due to the fact that inverse Ising inference is underconstrained: In fact the maximum likelihood procedure is neither underconstrained nor overconstrained, as the number of model parameters (fields and couplings) is exactly equal to the number of input constraints (univariate and bivariate marginals). 

Overfitting is prevented by regularization, which refers to corrections to account for finite sampling effects. Regularization can be implemented in various ways such as adding bias terms to the likelihood function, using early stopping, applying priors to model parameters, or adding noise to the inference procedure, and these strategies are often equivalent. Other studies using inverse Ising Inference have added $l_1$ or $l_2$ regularization terms to the log likelihood function $\ell$ which are functions of the coupling parameters of the Potts model, commonly a gauge-dependent $l_2$ term $R = \gamma \sum_{ij\alpha\beta} ( J^{ij}_{\alpha\beta})^2$ evaluated in the zero-mean gauge. Regularization comes at a cost of bias in the model, generally to weaken correlations. Regularization has been shown to improve contact prediction using Potts models when using other inference algorithms \cite{ZHvj,zm5F,OatS}. The use of regularization can introduce biases into the model predictions, which we investigate in results.

\subsection{Regularization}

In this study we regularize by applying a particular form of bias to the input bivariate marginals, chosen based on two principles. First, we wish to bias the observed bivariate marginals towards those of the independent model in order to help eliminate spurious correlations caused by finite sampling effects. Second, we would like to tune the strength of the bias such that the discrepancy between the observed marginal and biased marginal is equal to that expected due to sampling error, if one were to take a sample of size $N$ from the biased marginals. This should produce a regularized model which is still statistically consistent with the observed MSA.

This leads us to the following strategy. We compute the biased bivariate marginals as $\tilde{f}^{ij}_{\alpha\beta} = (1-\gamma^{ij}) \hat{f}^{ij}_{\alpha\beta} + \gamma^{ij} \hat{f}^i_\alpha \hat{f}^j_\beta$ for a choice of regularization strength $\gamma^{ij}$ which may differ for each position-pair, chosen as described further below, where $\hat{f}^{ij}_{\alpha\beta}$ refers to the marginals sampled from the MSA, $\tilde{f}^{ij}_{\alpha\beta}$ to the biased marginals, and $f^{ij}_{\alpha\beta}$ to the marginals of the Potts model. Varying $\gamma^{ij}$ from 0 to 1 interpolates between the MSA bivariate marginals and the corresponding site-independent bivariate marginals. This bias, \red{which behaves effectively like a pseudocount proportional to the univariate marginals}, preserves the univariate marginal constraints while weakening the (potentially spurious) correlations since $\tilde{C}^{ij}_{\alpha\beta}$ becomes 0 when $\gamma^{ij} = 1$.

This regularization strategy is equivalent to adding a regularization term to the likelihood function $R = - \sum_{ij} \gamma^{ij} \hat{X}^{ij}$ which biases the correlation energy terms defined above, and which is gauge-independent. Since $\frac{\partial R}{\partial J^{ij}_{\alpha_\beta}} = -\gamma^{ij} \hat{C}^{ij}_{\alpha\beta}$, using the fixed $\hat{C}^{ij}_{\alpha\beta}$ values from the dataset MSA, then the modified likelihood $\ell' = \ell + R$ is minimized (its gradient is 0) when $f^{ij}_{\alpha\beta} = \hat{f}^{ij}_{\alpha\beta} - \gamma^{ij} \hat{C}^{ij}_{\alpha\beta} =  (1-\gamma^{ij}) \hat{f}^{ij}_{\alpha\beta}  + \gamma^{ij} \hat{f}^i_\alpha \hat{f}^j_\alpha$, which is the bias formula used above. Thus, this form of regularization can be conveniently implemented as a simple preprocessing step to bias the bivariate frequencies, without the need to explicitly account for the regularization term in the quasi-Newton optimization procedure.

We choose the regularization strengths $\gamma^{ij}$ by finding the value such that the discrepancy between the observed marginals $\hat{f}^{ij}_{\alpha\beta}$ and the biased marginals $\tilde{f}^{ij}_{\alpha\beta}$ is equal to the expected discrepancy due to finite sampling. We measure this discrepancy using the ``Kullback-Leibler" (KL) divergence $\text{KL}(\hat{f}_{\alpha\beta}, \tilde{f}_{\alpha\beta}) = \sum_{\alpha_\beta} \hat{f}_{\alpha\beta} \log (\hat{f}_{\alpha\beta}/\tilde{f}_{\alpha\beta})$, which is a measure of the log-likelihood that a multinomial sample from the distribution $\tilde{f}_{\alpha\beta}$ would give the observed distribution $\hat{f}_{\alpha\beta}$. We choose the highest value $\gamma^{ij}$ such that the expected discrepancy $E[\text{KL}(F_{\alpha\beta}, \tilde{f}_{\alpha\beta})] \ge \text{KL}(\hat{f}_{\alpha\beta}, \tilde{f}_{\alpha\beta})$, where $F_{\alpha\beta}$ are sample marginals drawn from a multinomial distribution around $\tilde{f}_{\alpha\beta}$ with sample size $N$. This inequality can be solved numerically for $\gamma^{ij}$ by various means, and we show a fast and accurate approximation in appendix \ref{regstrength}.

\red{As an alternate regularization strategy, we also inferred models using $l_2$ regularization on the coupling parameters in the zero-mean gauge. However, we did not find a good heuristic for choosing the regularization strength. In \cite{OatS}, using a pseudolikelihood inference method, a constant strength of $\lambda = 0.01$ on the couplings was found to be appropriate for all families with varied $L$ and $N$, using a regularization term $R = \lambda \sum_{ij\alpha\beta} (J^{ij}_{\alpha\beta})^2$. In \cite{7TfN}, also using a pseudolikelihood implementation, a strength of $\lambda = 0.01 q (L-1)/2N$ was used, accounting for scaling factors in the likelihood in that study, which corresponds to $\lambda = 6.96/N$ for our kinase dataset and $\lambda = 0.8/N$ for our SH3 dataset. However in our inferences these values were too small and similar heuristics did not work consistently across our datasets.}

\subsection{Kinase and SH3 Reference Models}

For use in our \textit{in-silico} tests we infer ``reference" Potts models from natural protein MSA data obtained from Uniprot for the kinase and SH3 protein families. We pre-process the MSAs as described in previous publications \cite{xIOL}. First, given a set of sequences in a protein family we correct for phylogenetic relatedness. The Potts model assumes that each sequence in our dataset is drawn independently from the distribution $P(S)$, however in reality sequences from different organisms are phylogenetically related. We account for this in a standard way by downweighting sequences in proportion to the number of similar sequences, as described in \cite{xIOL}. We are investigating other approaches to account for phylogeny; this will be reported elsewhere. We also reduce the alphabet size $q$ from 21 residue types to fewer in a way which preserves the correlation structure of the MSA, as described previously \cite{xIOL}. \red{Finally, to avoid issues with unobserved residue counts of 0, we apply a very small pseudocount to the computed bivariate marginals for all models of $10^{-8}$.}

Our kinase reference model is inferred using 8149 effective sequences after phylogenetic weighting, starting from 127,113 raw sequences, and has $L=175$ and $q=8$. The SH3 reference model is inferred using 3412 effective sequences starting from 18,520 raw sequences, and has $L=41$ and $q=4$. 

Although these models are affected by finite-sampling error relative to any ``true" or empirical fitness landscape, this does not affect our \textit{in-silico} sampling tests below in which we treat these models as ``reference" or benchmark models and attempt to reproduce the reference model from finite MSAs generated from the reference models. The \textit{in-silico} tests are also unaffected by any potential biases caused by phylogenetic weighting or alphabet reduction since neither preprocessing step is used.

\subsection{Interaction Score}

To predict contacts using the Potts model we use a simple interaction score, a ``weighted" Frobenius Norm, which we have found improves contact prediction as described in a previous publication \cite{xIOL}. This is computed as $I^{ij} = \sqrt{\sum_{\alpha \beta} (w_{\alpha \beta}^{ij}J_{\alpha \beta}^{ij})^2}$ where $w_{\alpha \beta}^{ij} > 0$ are tunable weights, and is evaluated in a ``weighted" gauge with constraint $\sum_\alpha w_{\alpha \beta}^{ij} J_{\alpha \beta}^{ij} = 0$. In the case the weights $w_{\alpha \beta}^{ij} = 1$ this reproduces the unweighted Frobenius norm calculation. We use weights $w_{\alpha \beta}^{ij} = \sqrt{f_{\alpha \beta}^{ij}}$, in order to downweight the effect of rarely-seen mutants in the MSA.

\section{Results}

\subsection{Statistical Robustness of $E(S)$ as a Function of $N$, $L$, and $q$}

\begin{figure}
\centering
\includegraphics[width=\columnwidth]{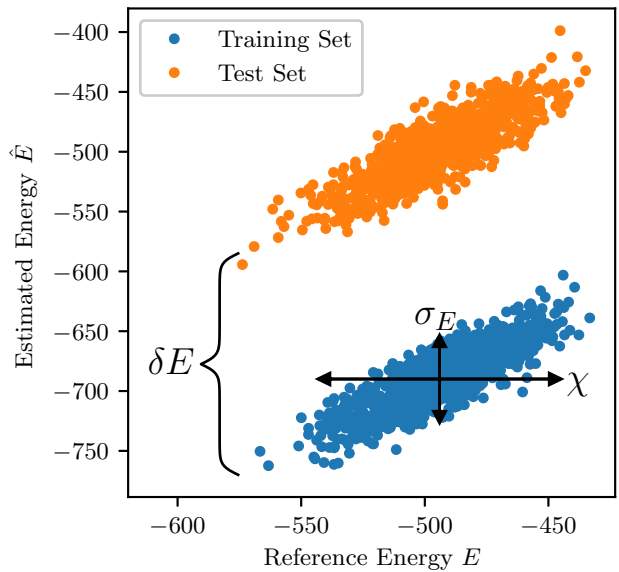}
\caption{Illustration of the ``signal to noise ratio" and the effects of finite sampling for the independent model. The SNR is the (squared) ratio of the $\chi$ to $\sigma_E$.  For this plot a ``reference" independent model was used to generate a ``test" MSA and a ``training" MSA of 1000 sequences each, and then a new independent model was parameterized using only the training MSA. $E(S)$ was then computed for the sequences of each MSA with both models. Finite sampling effects cause both the mean-squared-error $\sigma_E^2$, and an overfitting effect visible as a shift $\delta E$ of the estimated energies of the training MSA relative to those of the MSA.}
\label{fig:snr}
\end{figure}

Here we present a semi-quantitative discussion of the error in the statistical energy $E(S)$ of a sequence, the main quantity used to score and compare sequences, which is often interpreted as the sequence ``fitness", and which has been shown to predict experimental measures of fitness.

A measure of the statistical error in the Potts predicted energies for a set of sequences is the Pearson correlation coefficient $\rho(E, \hat{E})$ between the ``true" Potts statistical energy $E(S)$ according to a reference Potts model (which is unknown in the case of natural protein sequence datasets) to a reconstructed energy $\hat{E}(S)$ computed using a Potts model fit to a finite, limited depth MSA obtained by sampling from the reference model. The Pearson correlation coefficient is related to another useful quantity, the ``Signal to Noise" ratio (SNR), which is the ratio of the variance in statistical energies of sequences in the dataset, $\chi^2$, the ``signal", to the mean-squared-error in predicted statistical energies around their ``true" values, $\sigma_E^2$, the ``noise", or 
\begin{equation}
\text{SNR} = \frac{\chi^2}{\sigma_E^2}. 
\label{eq:SNR}
\end{equation}
The components of the SNR are illustrated in figure \ref{fig:snr}. 

If the reconstructed energies are modeled as the ``true" energies with added noise, i.e. $\hat{E}(S) = E(S) + \eta$ for noise $\langle \eta^2 \rangle = \sigma^2_E$, then $\rho(E, \hat{E}) = \sqrt{\text{SNR}/(\text{SNR}+1)}$. If the SNR or the Pearson correlation $\rho$ are small, the Potts model is unable to reliably distinguish high scoring sequences in the dataset from low scoring sequences. For an SNR less than 1 the typical energy difference between two sequences in the dataset will be smaller than the error, and their ranking according to the Potts model will be unreliable. This is important when using the Potts model to make fitness predictions.

Because of the mathematical challenges involved in analyzing analytically the Potts model's spin-glass behaviors, we illustrate the statistical effects of MSA depth using the independent model, a simpler but mathematically tractable model. We compute the expected $\chi^2$ and $\sigma_E^2$ and therefore the expected $\rho$. We then compare these results numerically with those of the full Potts model.

\subsection{The Noise Term $\sigma_E^2$}

Consider an MSA of $N$ sequences generated from an independent model, from which we estimate univariate frequencies $\hat{f}^i_\alpha$. The mean-squared-error in $\hat{f^i_\alpha}$ is $\sigma^2_{f^i_\alpha} = f^i_\alpha(1-f^i_\alpha)/N$ following a multinomial distribution. By propagation of error the mean-squared-error in the fields is $\sigma^2_{h^i_\alpha} \approx \frac{1- f^i_\alpha}{f^i_\alpha N}$, and we obtain the total mean-squared-error in the estimated energy of a sequence $S$ by summing these values for that sequence, $\sum_i^L \sigma^2_{h^i_{s_i}}$. Averaging over all sequences weighted by their probability, this gives
\begin{equation}
\sigma^2_E = \sum_S \sum_i^L \sigma^2_{h^i_{s_i}} = \sum_i^L \sum_\alpha^q f^i_\alpha \sigma^2_{h^i_\alpha} = \frac{L(q-1)}{N}.
\label{eq:sigmaE}
\end{equation}
This is the ``noise" part of the SNR, and corresponds to the vertical width illustrated in figure \ref{fig:snr}. It is equal to the number of independent model parameters $\theta^\text{Indep}$ divided by $N$. Intuitively, the statistical error in $E(S)$ increases with $L$ because $E(S)$ is a sum over $L$ parameters which each add a small amount of error, and it increases with $q$ because the average marginal, which is $\langle f^i_\alpha \rangle = 1/q$ by definition, decreases with $q$ and because fields corresponding to smaller marginals have greater error: The average mean-squared-error in field value is $\langle \sigma^2_{h^i_\alpha} \rangle = (q-1)/N$ which increases with $q$.

Absent strong correlated effects, the nature of this derivation suggests that the noise term for the Potts model can be estimated by replacing the number of parameters in the numerator with $\theta^\text{Potts}$, the number of independent Potts parameters. In practice correlated effects may cause deviations from this estimate, which we investigate numerically below.

We note that the approximation $\sigma^2_{h^i_\alpha} \approx \frac{1- f^i_\alpha}{f^i_\alpha N}$ used above is only valid if the sampled frequency $\hat{f}^i_\alpha$ is not small or 0. The case where the sample count is exactly 0 is particularly problematic as it leads to an inferred field $h^i_\alpha = \log(0) = -\infty$, meaning that the model predicts sequences with that residue can never be observed, which seems unreasonable. How to correct for the small-sample case depends on the user's prior expectations for the residue frequencies. For instance, one can add various forms of pseudocount \cite{Z9l8}. Because this is a somewhat subjective modeling choice, and because it does not affect our main results, we ignore small-sample corrections here although they are generally needed in practice.

\subsection{The Signal Magnitude $\chi^2$ and the SNR}

Next we compute $\chi^2$, the ``signal" part of the SNR. In the limit of large $L$ for the independent model one finds, using a saddle-point approximation, that the dataset sequence energy distribution is well approximated by a Gaussian distribution with variance $\chi^2 = \sum_{i}^L \chi_i^2$, with $\chi_i^2 =  \frac{1}{q} \sum_{\alpha} (h^i_\alpha)^2$ where the fields $h^i_\alpha$ are evaluated in the zero-mean gauge, as shown in appendix \ref{saddle}. $\chi_i^2$ can be thought of as a measure of the degree of conservation at position $i$ ranging from 0 to $\infty$. Unconserved positions with no sequence bias (all $h^i_\alpha = \log 1/q $ before gauge transformations) have $\chi^2_i = 0$, and highly conserved positions ($h^i_\alpha \rightarrow \infty$) will have very large $\chi^2_i$. We define the ``average per-site conservation" of the model $\langle \chi^2_i \rangle = \chi^2/L$, which should be independent of $L$. Combining these results we find the SNR of the independent model is given by
\begin{equation}
\text{SNR} = \frac{\chi^2}{\sigma^2_E} \sim \frac{N \langle \chi^2_i \rangle}{q-1}.
\label{eq:snr}
\end{equation}
The SNR for the independent model increases with the MSA sequence depth $N$ and the average per-site conservation $\langle \chi^2_i \rangle$, decreases with alphabet size $q$, and is independent of sequence length $L$.

\begin{figure}
\centering
\includegraphics[width=\columnwidth]{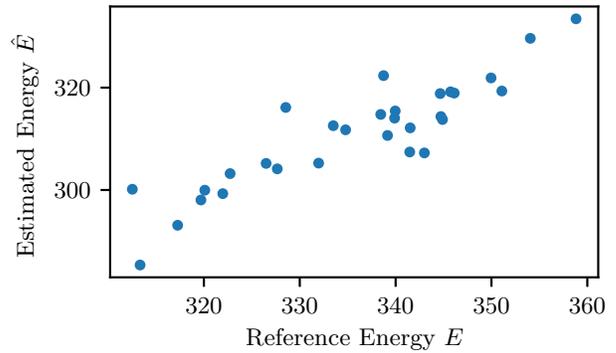}
\caption{\red{Example of $E(S)$ estimation for an independent model with 1600 parameters from an MSA with 30 sequences. The model has $L=200, q=8, N=30$ and $\langle \chi^2_i \rangle = 1.0$ for all $i$, and the fields are uniform random values scaled to give the correct $\langle \chi^2_i \rangle$. The scores for the 30 sequences have a correlation $\rho(E, \hat{E}) = 0.92$. }}
\label{fig:indep_smallN}
\end{figure}

This result shows that it is possible to accurately predict $E(S)$ even when the number of model parameters is much larger than the number of samples (the number of sequences). As an example, consider an independent model fit to a protein family MSA which is well-described by such a model, with $L=200$, $q=8$, and $\langle \chi^2_i \rangle = 1.0$, which appears to be typical of families in the Pfam database. Using equation \ref{eq:snr} one finds that only 30 sequences are needed to obtain a correlation of $\rho(E, \hat{E}) = 0.9$, while the model has 1600 parameters. This example is demonstrated numerically in figure \ref{fig:indep_smallN}.

In appendix \ref{saddle} we also show there that the Gaussian approximation only holds if $\langle \chi^2_i \rangle < 2 \log q$, which fails for highly conserved sequence datasets. For the kinase MSA, we find this inequality is $1.5 < 4.2$, and for the SH3 MSA we find $1.2 < 2.8$, so both MSAs have sufficient variation.

\subsection{Overfitting of $E(S)$ and $\delta E$}

Here we show how, for the independent model, overfitting results in a favorable energy shift of sequences in the training dataset (the MSA the model is parameterized with) relative to other sequences.

\begin{figure}
\centering
\includegraphics[width=\columnwidth]{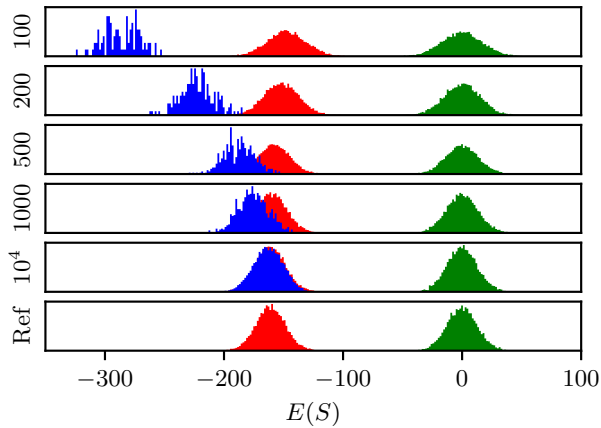}
\caption{Numerical tests of overfitting in the independent model. Each row corresponds to an independent model fit to a ``training" MSA dataset with different MSA depth $N$, generated from a reference independent model. The training MSAs have $L = 1000$, $q=16$, and $\langle \chi_i^2 \rangle = 0.16$. A pseudocount of $1/N$ is used to avoid issues with unsampled residues. The green distribution shows estimated energies of ``random" sequences with equal residue probabilities, the blue distribution shows energies of training MSA, and the red distribution are energies of a ``test" MSA independently generated from the reference model. The models are evaluated in the zero-mean gauge.}
\label{fig:independent_test}
\end{figure}

When a single sequence is added to an MSA of size $N-1$ the estimated site-frequencies for the residues $i,\alpha$ in that sequence are increased to ${^+\hat{f}^i_\alpha} = ((N-1)\hat{f}^i_\alpha+1)/N$ the rest decrease to ${^-\hat{f}^i_\alpha} = ((N-1)\hat{f}^i_\alpha)/N$, where $\hat{f}^i_\alpha$ is the original sampled marginal. The prevalence of the added sequence in the new model is then $P^+(s) = \prod_i {^+\hat{f}^i_{s_i}}$, while previously it was $P(s) = \prod_i f^i_{s_i}$ on average. The ratio of these prevalences, averaging over all possibilities for the added sequence, is $\sum_S P(S) \frac{P^+(S)}{P(S)} = (\frac{N+q-1}{N})^L \approx e^{\frac{L(q-1)}{N}} =  e^{\sigma_E^2}$, in the large $L$ limit. This corresponds to a relative statistical energy change of 
\begin{equation}
\delta E = \sigma^2_E
\label{eq:deltaE}
\end{equation}
to a sequence when it is added to the training MSA. In other words the predicted energies for sequences used to train the model will be underestimated (i.e, their favorability is overestimated) by an amount $\delta E$ which decreases with $N$. This is typical of the effect of overfitting in other contexts. This overfitting effect is confirmed using numerical tests in figure \ref{fig:independent_test}.

This suggests that when the Potts energy is used to score sequences, care should be taken if the sequences to be scored contain both sequences from the training set as well as other sequences, as there may be an energy shift between the two types. In our numerical tests below we investigate whether this affects common applications of the Potts model such as predicting statistical energy changes $\Delta E$ caused by mutation in a sequence in the training set.

\subsection{\textit{In Silico} Tests of Potts Model Robustness in $E(S)$ as a Function of N}

Next we numerically test the behavior of the Potts model inference for different MSA depths using an \textit{in-silico} procedure. We use Potts models parametrized for the protein-kinase and SH3 domains using Uniprot sequence data as reference models, as described in methods. We then generate new MSAs from these reference models, of depths of 256 to 16384 sequences, from which we infer new models. For each generated MSA, we fit both an unregularized and a regularized model. 

\red{The reference models used in these \textit{in-silico} tests are derived from real protein-family MSAs, and therefore have mutational correlation patterns close to those of the real SH3 and kinase protein families albeit with some errors due to finite-sampling effects. Both families we study have very deep MSAs and we expect small statistical error due to finite sampling of the MSA. We expect that the strength of the correlations and the degree of sparsity of the interaction network of our reference models are representative of protein family MSAs like those collected in the Pfam database. It is important to keep in mind that other types of data such as neuron spike-trains may have different properties, e.g. they may behave more or less ``critically"\cite{rSDu}, or have less sparse interaction network, which may make the inference problem more or less difficult. Our numerical tests of finite-sampling error specifically use protein-family-like data, although we expect our results are more general.}

\begin{figure}
\centering
\includegraphics[width=\columnwidth]{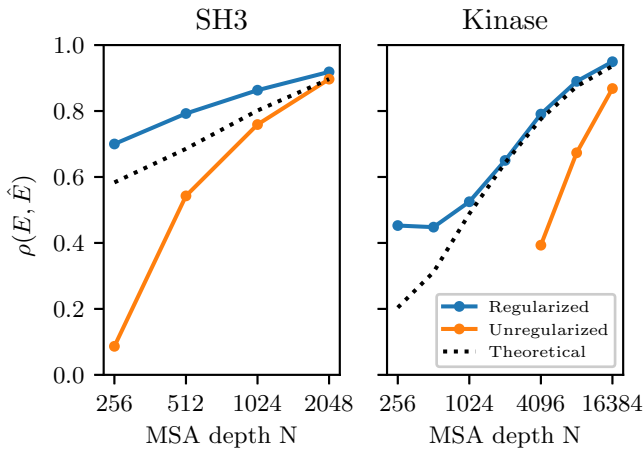}
\caption{Accuracy in Statistical Energy Predictions, measured by the Pearson correlation coefficient $\rho(E, \hat{E})$ between the reference energies $E$ and the inferred energy $\hat{E}$ for the \textit{in-silico} MSA, as a function of MSA depth $N$, for the SH3 and kinase domain for both regularized an unregularized inference. The theoretical curve is computed using equation \ref{eq:potts_rho}}
\label{fig:pearson_depth}
\end{figure}

For each \textit{in-silico} model, after confirming convergence of the inference procedure, we evaluate its predictive accuracy by computing the Pearson correlation $\rho(E, \hat{E})$ between the predicted Potts statistical energies and those computed using the reference models, for the sequences used to train the new models. We also compute the expected $\rho$ using equations \ref{eq:SNR} and \ref{eq:sigmaE} modified for the Potts model, giving
\begin{equation}
\rho(E, \hat{E})^2 = \frac{\chi^2}{\chi^2 + \frac{\theta^\text{Potts}}{N}}
\label{eq:potts_rho}
\end{equation}
where $\theta^\text{Potts}$ is the number of Potts model parameters described above and $\chi^2$ is estimated from the variance in inferred sequence energies. Results are shown in figure \ref{fig:pearson_depth}.

We find that the unregularized models which are fit to smaller MSAs are overfit, with two clear symptoms. First, for the kinase MSAs, which have a greater number of parameters because of their larger $L$ and $q$, the unregularized MCMC inference procedure fails to converge in reasonable time for small MSAs with $N \le 2048$. The behavior is consistent with the Potts model becoming ``ill conditioned", which is a predicted consequence of finite sampling error \cite{wa7p}. For these small MSAs, as the Potts parameters are successively updated we find that the MCMC sampling step takes longer and longer to equilibrate, eventually slowing to a standstill in which MCMC replicas appear to be trapped in local wells in a rugged landscape, and the auto-correlation time described in methods diverges. Second, even for the unregularized models which we were able to converge, which are the kinase models for $N \ge 4096$ and the SH3 models, we find that after a finite number of parameter update steps the model error begins to increase (see appendix \ref{convergence}). This is behavior typical of overfitting. This effect decreases for larger $N$, and we find that for $N=16384$ for kinase, and for $N=2048$ for SH3, these overfitting effects are minimal. The effects of overfitting can be mitigated through regularization, and we find that for our regularized inference the autocorrelation time always decreases rapidly and the model error does not increase much after many iterations. The regularized model error nevertheless increases slightly from its minimum value after many iterations, suggesting it is still slightly overfit.

For the models which converged we find, as expected, that the model error decreases with $N$, as shown in figure \ref{fig:pearson_depth}. For both kinase and SH3, the unregularized models have more error than expected based on our theoretical analysis. Regularization significantly reduces the error, especially for small MSAs. For equal $N$ we see that the SH3 model has less error than the kinase model, as expected since the SH3 model has smaller $L$ and $q$. For both protein families we find that the theoretical result is better than that of the unregularized model, perhaps because of correlated effects, but that with regularization the model outperforms our theoretical expectation based on the error analysis of the independent model. For the largest $N$ of 16384 for kinase and 2048 for SH3, the unregularized models perform almost as well as the regularized model, again suggesting that regularization is largely unnecessary with this many sequences even though this depth is much smaller than the number of parameters of the models, of 747250 and 14883 respectively. This is a further demonstration that the effect of overfitting is best estimated from the signal to noise ratio, and not directly from the number of parameters of the model. \red{We note that even for large MSAs some form of regularization of may still be necessary to prevent some model parameters from becoming infinite in the case of unobserved residue-pairings, for instance by addition of a small pseudocount as discussed above in the derivation of $\sigma^2_E$. }

\begin{figure}
\centering
\includegraphics[width=\columnwidth]{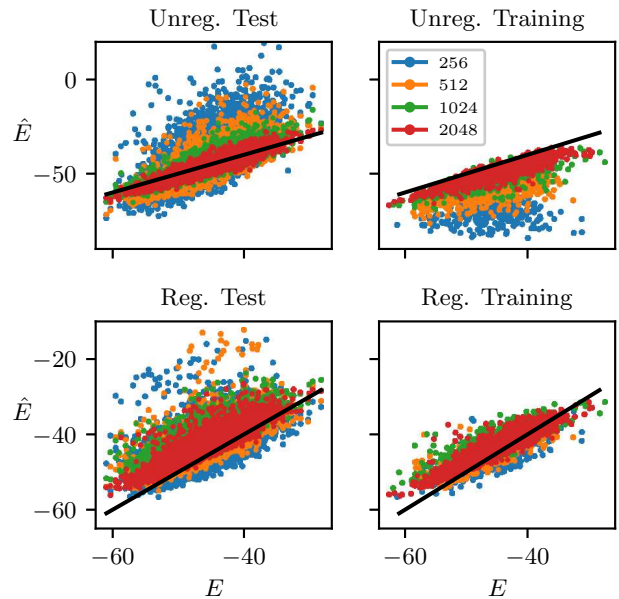}
\caption{Comparison of statistical Energy Predictions relative the reference values for the SH3 dataset, for models fit to different MSA sizes, for both test and training MSAs, with and without regularization. All models are evaluated in the zero-mean gauge.}
\label{fig:SH3_E_scatter}
\end{figure}

We also examine the $\delta E$ shift, or average change in the dataset sequence probabilities, caused by overfitting. We can estimate $\delta E$ as the difference in mean energy of sequences in the MSA used to train the model (a ``training set") and a separate set of sequences generated by the reference model (the ``test set"). For the converged unregularized models we find a negative $\delta E$ shift consistent with our expectation from the independent model, which decreases with $N$, as seen in figures \ref{fig:SH3_E_scatter}. For the regularized models we also observe a $\delta E$ shift, but it is positive and invariant with $N$. The existence of these $\delta E$ shifts has implications for applications of the Potts model which depend on the absolute probability of sequences in the dataset. For instance, the energy average has been used to estimate the size of the evolvable sequence space \cite{XwuF}, and the energy gap between ``random" sequences and the dataset sequences has been used to estimate the ``design temperature" of the Random Energy Model of protein evolution\cite{wiT3}. The fact that the inferred $\delta E$ depends on the choice of regularization or on the MSA depth suggests such computations should be calibrated by other means, for instance by referring to experimental melting temperature as in \cite{wiT3}.

\subsection{Mutation Effect Predictions}

\begin{figure}
\centering
\includegraphics[width=\columnwidth]{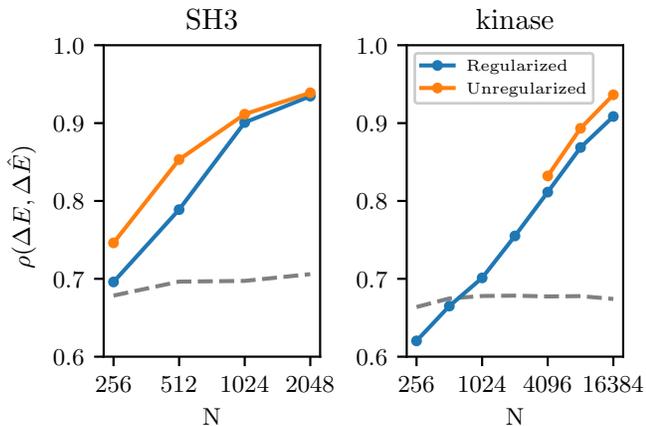}
\caption{Accuracy of point-mutation effect predictions as a function of MSA depth. This is measured by the Pearson correlation in mutation effect $\Delta E$ for all possible mutations to a set of 100 sequences generated from the reference model. The dashed line is the Pearson correlation for mutation effects predicted by an independent model fit to the univariate marginals of each \textit{in-silico} MSA, with a pseudocount of 0.5 counts.}
\label{fig:mutation_rho}
\end{figure}

A common application of Potts statistical energies is in predicting the effect of a mutation to a sequence, by computing the change in statistical energy $\Delta E$ after a small number of positions have been mutated. The Potts model has been shown in many cases to predict mutation effects quite accurately \cite{7TfN}, and importantly the correlated nature of the Potts model makes these predictions ``background dependent", meaning that same mutation in two different sequences can have a different effect. Above we predicted that due to overfitting the unregularized Potts model can score sequences it was inferred with more favorably than other sequences, which could conceivably affect mutation effect predictions involving a mutation from a sequence in the training set to one not in the training dataset.

To test the effects of MSA size and overfitting on mutation effect predictions, we generated a set of 100 sequences from the kinase and SH3 reference models, and computed the change $\Delta E$ caused by all point-mutations to each sequence using the reference models and then again using each of the \textit{in-silico} models, and measured the discrepancy using the Pearson correlation $\rho(\Delta E, \Delta \hat{E})$. We find that the accuracy of point-mutation predictions decreases with $N$, but much less quickly than that of the energy of entire sequences $E(S)$ (figure \ref{fig:pearson_depth}), and even for our smallest MSAs of 256 sequences we find a correlation of $0.7$ for the SH3 model and $0.6$ for the kinase model (figure \ref{fig:mutation_rho}). With 16384 effective sequences for the kinase family we find a correlation of $\sim 0.9$ with the reference. Previously reported values of the correlation between Potts mutation effect predictions and experimental measures of fitness are in the range 0.5 to 0.8 for MSAs with fewer than 10,000 effective sequences \cite{7TfN}. 

We also find that for the smallest MSAs an independent model performs nearly as well or better than the Potts model for point-mutation-prediction (dashed line in figure \ref{fig:mutation_rho}). This suggests that for very small MSAs the benefits of the correlated information in the Potts model are diminished by its increased statistical error and poorer signal-to-noise ratio. \red{Indeed, in \cite{7TfN} it was found that the independent model performed comparably or better than the Potts model in $\Delta E$ predictions for some datasets. In contrast, when predicting full statistical energy $E(S)$ as in figure \ref{fig:pearson_depth} the independent model performs very poorly compared to the Potts model even with very small MSAs, giving a $\rho(E, \hat{E})$ of 0.4 for SH3 and -0.1 for kinase when fit to the reference model's univariate marginals. These results suggest that correlated effects are less important when predicting single-mutant $\Delta E$ values, and that for small MSAs the Potts model behaves roughly like the independent model in this application.}

\begin{figure}
\centering
\includegraphics[width=\columnwidth]{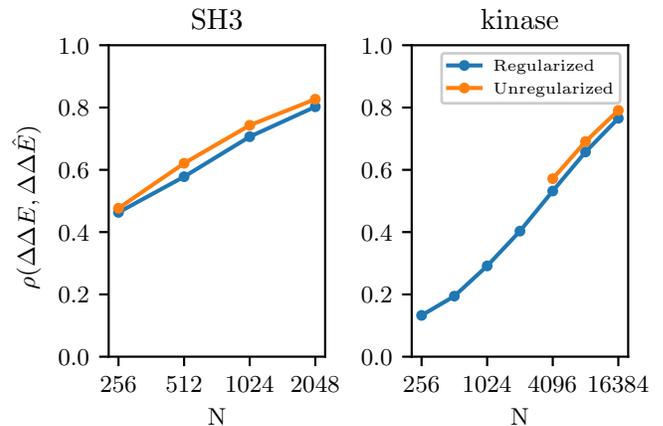}
\caption{\red{Accuracy of ``double mutant cycle" predictions of epistasis as a function of MSA depth. This is measured by the Pearson correlation in mutation effect predictions $\Delta \Delta E$ for all possible double mutants to a set of sequences generated from the reference model.}}
\label{fig:ddE}
\end{figure}

\red{In addition to point-mutation predictions, we examine double-mutant predictions of the form commonly used to test for epistasis (non-additivity of fitness effects) in experimental ``double mutant cycles"\cite{kgPd,Ado2}. Here, deviations from an independent model are tested using the quantity $\Delta \Delta E^{ij}_{\alpha\beta}(S) = \Delta E^{ij}_{\alpha\beta}(S) - \Delta E^{i}_{\alpha}(S) - \Delta E^{j}_{\beta}(S)$ where the subscripts indicate which positions are mutated. This is the difference in mutation-effect between a double mutant and the sum of the two corresponding single mutants. The independent model cannot predict these values as it gives $\Delta \Delta E^{ij}_{\alpha\beta}(S) = 0$ by definition. In contrast for the Potts model one obtains the gauge-invariant result $\Delta \Delta E^{ij}_{\alpha\beta}(S) = -J^{ij}_{\alpha\beta} + J^{ij}_{\alpha s_j} + J^{ij}_{s_i \beta} - J^{ij}_{s_i s_j}$. We test the model's ability to reconstruct these values by generating sequences from the reference models and then comparing the predicted and reference $\Delta \Delta E$ values for all possible double-mutants to each sequence scored as $\rho(\Delta \Delta E, \Delta \Delta \hat{E})$, and the result is shown in figure \ref{fig:ddE}. We find that the quality of the $\Delta \Delta E$ prediction degrades much more rapidly with $N$ than single-mutant $\Delta E$ prediction, showing that deep MSAs are very important to capture the correlated effects that are probed by double mutant cycles, which depend on accurate predictions of $\Delta \Delta E$.}

\subsection{Contact Prediction}

Using the same \textit{in-silico} datasets we test the accuracy of contact prediction as a function of $N$. Because there is no unique mapping from the Potts model parameters to contact predictions, many different mappings have been suggested. The most straightforward methods compute an ``interaction score" for each position-pair $i,j$ which is a simple function of the coupling parameters and marginals only involving those positions. These include the ``Direct Information", the Frobenius norm, and the weighted Frobenius norm. Typically some fraction of the highest scoring pairs, for instance the top $L$, are chosen as predicted contacts. Recently, more advanced machine learning algorithms have been used, trained using external structural data, to find more complex mappings from the coupling parameters to contact predictions, which have shown increased predictive accuracy\cite{Sufl,XPxT,B0Tq}. 

\begin{figure*}
\centering
\includegraphics[width=\textwidth]{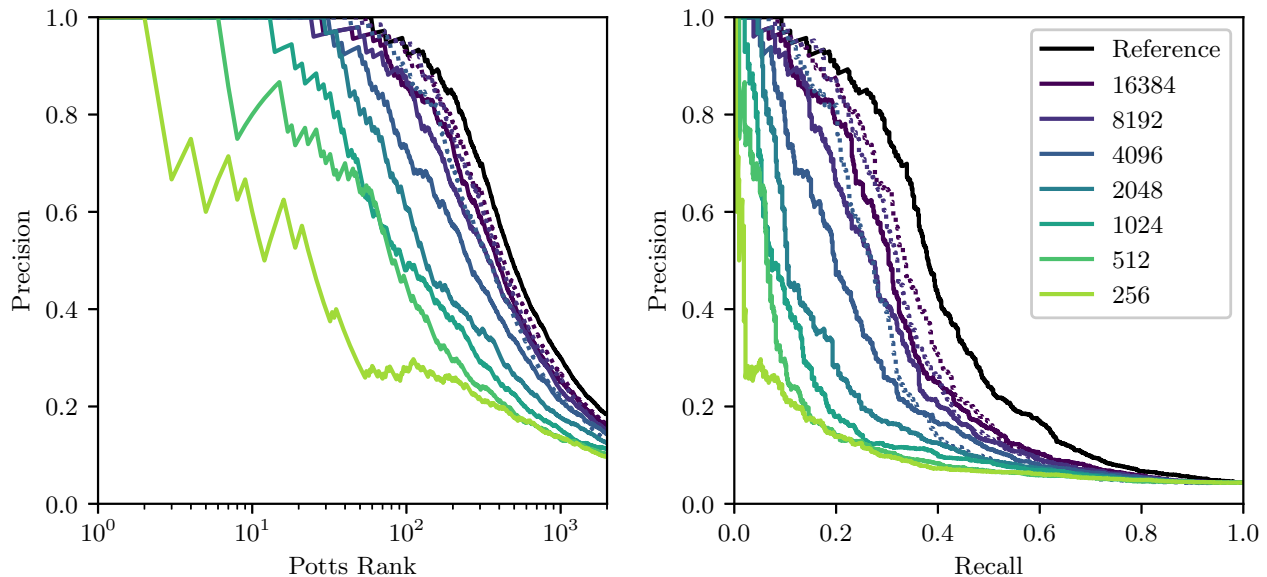}
\caption{Contact prediction as a function of MSA depth $N$, for position pairs with $|i-j| > 4$. Left: Fraction of correctly predicted contacts (precision, or $\text{TP}/\text{rank}$) versus position-pair rank, ordered by Potts interaction score, for the original Potts model and for derived Potts models fit to smaller MSA depths. The solid line corresponds to the regularized models, and the dotted lines correspond to unregularized models. Right: Precision-recall plot for the same models, by varying the rank cutoff. ``Precision" is computed as $\text{TP}/(\text{TP} + \text{FP})$, and ``recall" as $\text{TP}/(\text{TP} + \text{FN})$. $\text{TP}$ (``true positive") is the number of correctly predicted contacts, $\text{FP}$ (``false positive") the number of contacts predicted but not present in crystal structure, and $\text{FN}$ (``false negative") the number of crystal contacts not predicted by the model.}
\label{fig:contacts_deep}
\end{figure*}

Here we focus on the effects of finite sampling on the (weighted) Frobenius norm. We begin by analyzing the baseline contact predictions of the reference kinase Potts model, \red{which will serve as an upper limit to the performance in our \textit{in-silico} models.} The existence of extensive crystallographic data on the kinase family in the Protein Data Bank (PDB) \cite{pREo} makes it especially well suited for testing contact prediction, as it has been shown that Potts interactions can correspond to transient contacts across multiple functional conformations \cite{wRvM,NBbG,8K1h,I7GI} which we can detect using the large PDB dataset. To define reference ``true contacts" we average over 3000 kinase structures in the PDB. We count a contact between positions-pair $(i, j)$ if the residues have a heavy atom pair within 6 Angstroms in at least 20\% of the PDB structures determined in \cite{8K1h}, giving us a set of 1180 total contacts for the kinase family. Using the ``weighted Frobenius Norm" interaction score we find that 80\% of the top 511 most strongly interacting position-pairs predicted by the model are contacts in the PDB. Limiting our analysis to position-pairs which are distant in sequence, with $|i-j| > 4$ as is typical in tests of contact prediction, we find that 80\% of the 176 most strongly scored of these pairs are PDB contacts, out of 637 relevant contacts (precision=0.8, recall=0.22), as illustrated by the black line in figure \ref{fig:contacts_deep}.

\red{Next we use the \textit{in-silico} models for contact prediction.  We note that the \textit{in-silico} models should not perform better than the reference model since any discrepancy between the crystal contacts and the contacts predicted by the reference model will be inherited in the \textit{in-silico} models, and so the reference model result represents the maximum possible result for the \textit{in-silico} models except for small statistical variations. The \textit{in-silico} tests measure how finite-sampling error further degrades the result from our reference model.}

As MSA depth $N$ decreases for the \textit{in-silico} models we find that contact prediction accuracy decreases, as illustrated in figure \ref{fig:contacts_deep}. We see a more minor decrease in contact prediction accuracy from 16384 sequences to 4096 sequences, and then a more dramatic drop from 4096 to 256 sequences, but even the smallest models are able to predict some contacts. These results suggest that for the purpose of contact prediction, compared to statistical energy predictions, it is more important to have deeper MSAs. Our unregularized models, when converged, also appear to have very similar performance as the regularized models for contact prediction.

\red{We note that discrepancy between contacts predicted using the reference model and crystal contacts may not be due to biases in the Potts
model, but rather in the definition of crystal contacts or in the interaction
scoring function which is used as a proxy for contacts.} A series of previous studies has examined contact prediction
by different crystal contact definitions and scoring methods including scorings
determined by machine-learning\cite{Rve7,Sufl,TzRy}. These studies measure the
overall prediction accuracy using the precision (y-axis in figure
\ref{fig:contacts_deep}) for the $C$ top-ranked pairs according to the model
(corresponding to a value on the x-axis in figure \ref{fig:contacts_deep},
left), where $C$ has different values in different studies, typically $C$ is
$L$, $2L$, $2.5L$ \cite{TzRy} or $n_\text{contacts}/2$ where
$n_\text{contacts}$ is the number of contacts observed in the reference crystal
structure\cite{Sufl}. The definition of a structural contact also differs. All studies
exclude position-pairs where $|i-j| \le 4$, but some studies use the distance
between $\operatorname{C-\beta}$ atoms while others use the closest heavy-atom distance, and
the distance cutoff varies from 10 \AA\ to 6 \AA. As discussed in \cite{Rve7},
increasing the distance cutoff will always increase the precision and will
inflate the apparent performance, and for this reason we also plot recall in figure
\ref{fig:contacts_deep}. For comparison with previous studies, using the
weighted Frobenius Norm and $C = L = 175$, we get precisions of 0.76 and 0.86
respectively for the 8\AA\ $\operatorname{C-\beta}$ and 6\AA\ heavy-atom contact definitions.
For $C = n_\text{contacts}/2 = 318$ this gives precisions of 0.71 and 0.68
respectively. In previous studies, an ``Average Product Correction" has been
applied to the Frobenius Norm scores, we find this decreases the precision, for
instance to 0.46 and 0.41 respectively for $C = n_\text{contacts}/2$.

\subsection{Literature Review of Model Sizes and Estimated Statistical Errors}

\begin{figure}
\centering
\includegraphics[width=\columnwidth]{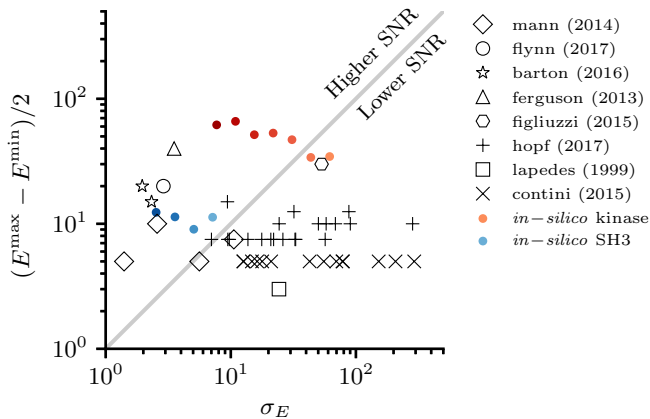}
\caption{Analysis of Potts model parameters in published literature. The estimated ``noise" in $E(S)$, $\sigma_E$, is computed by dividing the number of Potts parameters by the MSA depth, as in formula \ref{eq:sigmaE}, and is compared to the energy difference between the highest and lowest sequence energies plotted in the publication, reflecting the ``signal". Red: Kinase \textit{in-silico} regularized models in this study for N=16384 (darkest) to 256. Blue: SH3 \textit{in-silico} regularized models in this study for N=2048 (darkest) to 256. Without regularization and according to the naive results suggested by the independent model, Potts models below the diagonal line should have difficulty discriminating between the highest and lowest scoring sequences in the dataset. The cited studies are: Mann \cite{bgwj}, Flynn \cite{IiTK}, Barton \cite{VQDL}, Ferguson \cite{Cmda}, Figliuzzi \cite{W5Kp}, Hopf \cite{7TfN}, Lapedes \cite{3LtB} and Contini \cite{BRO2}. }
\label{fig:SNR}
\end{figure}

Our analysis of the independent model illustrates how model quality depends on the SNR, which is the (squared) ratio of the ``noise" due to statistical error $\sigma_E$, which depends on $N$ and the number of model parameters, to the ``signal" strength $\chi$, which depends on $L$ and degree of conservation. Our numerical results for $\rho(E,\hat{E})$ as a function of MSA depth suggests that substituting the Potts model's number of parameters $\theta^\text{Potts}$ into equation \ref{eq:sigmaE} gives an estimate of $\sigma_E$ for the Potts model (see figure \ref{fig:pearson_depth}). This gives us a way to estimate the statistical error of models published in literature given the published values of $L$, $q$, $N$ and $\chi$.

We have collected model parameters from a number of previous studies in literature. For each model we estimate $\sigma_E$ given the published $N$, $q$ and $L$. $\chi$, which measures sequence conservation, is not typically published, but we estimate it from the range of reported $E(S)$ values for each model, as $(E^\text{max} - E^\text{min})/2$. This will be an overestimate of $\chi$, as $\chi$ represents the standard deviation in energy values rather than the range, but nevertheless should roughly correspond. These results are summarized in figure \ref{fig:SNR}.

\red{ Many models, including most of those used in this study, are in the estimated ``higher SNR" region. Some models are below the diagonal, suggesting that greater MSA sequence depths could improve these models. Many of these low-SNR models were used mainly for predicting point-mutation $\Delta E$ or on small numbers of mutations (\cite{7TfN,BRO2}), which we showed above numerically can still be accurately predicted even with low SNR. However, in these cases the Potts model may not outperform an independent model, and indeed in \cite{7TfN} a number of these low-SNR models were compared to an independent model and found to have similar predictive accuracy in $\Delta E$. Additionally, we note that the estimates of $\chi$ in the y-axis in figure \ref{fig:SNR} are only rough estimates, and also that details of the inference procedure such as the regularization strategy can also help improve the predictive ability of the model past our expectations based on analysis of the independent model. This means that some models which appear below the diagonal may have greater statistical power than illustrated.}

\section{Conclusions}

Finite sampling error and overfitting play an important role in all inference problems, and Potts models are no exception. In this study we examined how finite sampling, which corresponds to MSA depth, affects common uses of Potts models for protein sequence analysis, which are: The prediction of individual sequence total statistical energies (often interpreted as fitnesses or in some cases as predictors of folding free energy), the prediction of the fitness effects of mutations to a sequence, the prediction of double-mutant epistatis, and the prediction of directly statistically dependent pairs of positions in the protein family, and their correspondence to contacts in 3D protein structure. 

Overfitting is ultimately due to finite-sampling statistical error in the bivariate marginals which serve as input variables to the model inference procedure, and which are estimated from as MSA with a finite number of sequences $N$. We discussed in a semi-quantitative way how this statistical error can affect Potts model predictions based on analysis of how the error depends on the Signal to Noise ratio (SNR) in a simplified model. The effects of finite sampling are a function of the dataset MSA length $L$, alphabet size $q$, MSA depth (number of sequences) $N$, and the degree of conservation of columns of the MSA, as measured by $\langle \chi^2_i\rangle$. From these quantities we can roughly estimate the expected Pearson correlation $\rho(E,\hat{E})$ between the ``true" sequence statistical energies and those predicted from a finite depth MSA. We arrived at these results using an independent model framework, but increasing the number of parameters from $\theta^\text{Indep}$ to $\theta^\text{Potts}$, and so the existence of strongly statistically dependent correlated networks among the positions of the MSA may cause deviations from these predictions. Nevertheless, for the kinase and SH3 models we studied we found numerically that it gives a reasonably good approximation.

We find that the different types of predictions based on Potts models of protein covariation are differently affected by finite sampling error and regularization. Predictions of the effect of point-mutations to a sequence $\Delta E$ are the most robust, while predictions of total statistical energies $E(S)$ decrease more rapidly in accuracy as a function of $N$. Contact prediction precision-recall curves for the smallest MSA depths we tested depend strongly on MSA depth and are poor, though we are able to predict tens of contacts with high confidence even with 256 sequences for the kinase model. Using our regularization strategy and MCMC inference procedure, we found that predictions of full sequences energies $E(S)$ are most improved by regularization, while mutation-effects predictions and contact predictions are slightly better with unregularized models, for the large MSAs which are possible to fit without regularization besides a very small pseudocount. Additionally, we find that in unregularized models fit to large MSAs the effects of overfitting can be negligible even in cases where the number of sequences (samples) is many orders of magnitude smaller than the number of Potts parameters, because overfitting effects are best understood in terms of the SNR and not directly from the number of Potts model parameters.

We also found that finite sampling and overfitting can cause an energy shift $\delta E$ in the predicted sequence energies $E(S)$ for sequences in the MSA used to parameterize the model. This shift is affected by regularization. This energy shift may be important to be aware of when performing computation which depend on the absolute value of the energy, or on the energy difference between sequences used to train the model and other sequences. Similarly, our observation of a divergent autocorrelation time when generating simulated MSAs by MCMC for unregularized models suggests that the ruggedness of the inferred energy landscape depends on the inference procedure and choice of regularization. Such computations should be calibrated by external means. 

In this study we have examined the contribution of finite sampling to the error in Potts model predictions, but there are other potential sources of error. These include biases in the input MSA, for instance due to errors in the sequence search and alignment procedure, or because of violations of the Potts modelling assumptions that the sequences have evolved independently over the same fitness landscape, for instance due to phylogeny, mutational biases, or variation in selective pressures over time and environment. In addition there are many assumptions that must be made to connect the various kinds of experimental measurements of fitness with Potts model predictions, even assuming no errors in the Potts model of the kind that are the focus of this work. We hope the results presented here clarify the baseline statistical power and limitations of Potts models of protein covariation, on which further understanding of the relationship between Potts models and the evolution and structure of proteins can be built.

\begin{acknowledgments}
This work has been supported by grants from the National Institutes of Health (R01-GM30580, U54-GM103368). We also acknowledge useful discussions with Jonah McDevitt and William F. Flynn, and assistance for literature search from William F. Flynn.
\end{acknowledgments}

\appendix
\section{\label{saddle} Saddle-Point Approximation for $\chi^2$}

Here we derive the distribution of energies of the independent model building on \cite{GaDX}. The value of $\chi^2$ can be estimated given a set of fields $h^i_\alpha$ of the independent model using a saddle-point approximation in the limit of large $L$. To do this we first compute the ``neutral" or ``background" distribution of sequence energies $\Omega(E)$, showing to good approximation it is Gaussian with variance $\chi^2$. This distribution plays the role of the (normalized) ``density of states" in statistical mechanics, and may be written as
\begin{equation}
\Omega(E) = \sum_S^{q^L} P_0(S) \delta(E - E(S))
\end{equation}
where $P_0(S)$ is the ``background" probability of the sequence $S$, which in this study we approximate is uniform $P_0(S) = \frac{1}{q^L}$. Using the method of steepest descent (saddle-point approximation) we expand the delta function using its Fourier transform, giving
\begin{align}
\Omega(E) &= \frac{1}{2 \pi} \sum_S P_0(S) \int_{-\infty}^{\infty} e^{i k (E - E(S))} dk \\
&= \frac{1}{2 \pi i} \int_{-i \infty}^{i \infty} e^{\beta E + \text{ln}[Z]} d \beta 
\end{align}
with $\beta = i k$ and a partition function $Z(\beta) = \sum_S P_0(S) e^{-\beta E(S)}$. Expanding the exponent in the integral around its maximum at $\beta^*$ along the path of integration going through a minimum along the real axis, we identify $\beta^*$ by approximating the exponent using a high-temperature expansion around $\beta = 0$, 
\begin{align}
\beta E + \text{ln}[Z(\beta)] \approx & \beta E + \beta \frac{\partial \text{ln}[Z]}{\partial \beta}\vert_{\beta = 0}  + \frac{ {\beta}^2 }{2} \frac{\partial^2 \text{ln}[Z]}{\partial {\beta}^2} \vert_{\beta = 0}  \label{eq:aaa} \\ 
\equiv& \beta ( E - \langle E \rangle) + \frac{{\beta}^2}{2} \chi^2
 \label{eq:hightempB}\end{align}
with
\begin{align}
\frac{\partial \text{ln}Z}{\partial \beta}\vert_{\beta = 0} &= -\sum_{i}^L \sum_\alpha^q  \frac{1}{q} h^i_\alpha = -\sum_{i}^L \langle h\rangle^i \equiv -\langle E \rangle \\
\frac{\partial^2 \text{ln}Z}{\partial \beta^2} \vert_{\beta = 0} &= \sum_{i}^L \sum_{\alpha} \frac{1}{q} (h_\alpha^i - \langle h\rangle^i )^2 \equiv \chi^2 .
\end{align}

The maximum $\beta^*$ at which the first derivative of equation \ref{eq:hightempB} is 0 is then
\begin{equation}
\beta^* = -\frac{ E - \langle E \rangle}{\chi^2}.
\end{equation}
This approximation in the region near $\beta = 0$ is justified as long as $\beta^*$ is close to 0, when $\langle E \rangle - E \ll \chi^2$.

To complete the saddle-point analysis, we expand the exponent $\beta E + \ln[Z]$ again but around $\beta^*$, where the first derivative should be zero, giving
\begin{align}
\Omega(E) &=  \frac{1}{2 \pi i} \int_{-i \infty}^{i \infty} e^{\beta^* E + \text{ln}[Z(\beta^* )] + \frac{1}{2} (\beta - \beta^*)^2 \frac{\partial^2 \text{ln}Z}{\partial \beta^2} \vert_{\beta^*}} d \beta \\
&\propto  e^{\beta^* E + \text{ln}[Z(\beta^* )]} \frac{1}{2 \chi^2 } \\
&= \frac{1}{\sqrt{2 \pi \chi^2}} e^{-\frac{(E - \langle E \rangle)^2}{2 \chi^2} }
\end{align}
and so the density of states is a Gaussian distribution with mean $\langle E \rangle$ and variance $\chi^2$. For the independent model, as discussed in the main text we can always transform to the ``zero-mean" gauge in which $\langle E \rangle = 0$ and $\chi^2 = \sum_i^L \frac{1}{q} \sum_\alpha^q (h^i_\alpha)^2$, using the transform $h^i_\alpha - \frac{1}{q} \sum_\gamma^q h^i_\gamma \rightarrow h^i_\alpha$.

The distribution of ``evolved" sequence energies $P(E)$, i.e. of sequences generated by the independent model with probability $P(S)$, can then be written
\begin{equation}
P(E) \propto \Omega(E) e^{-E}
\end{equation}
using the Potts probability $P(S) \propto e^{-E(S)}$, and some algebra shows that this is also a Gaussian distribution, with mean $\langle E \rangle - \chi^2$ and variance $\chi^2$.

This approximation will break down when the ``evolved" sequences would have energies outside the range of validity of the Gaussian approximation, and we can estimate when this occurs. The sequence space has a size of $q^L$ sequences, so the density of states may be estimated as $q^L \Omega(E)$. The Gaussian approximation will break down for energies $E$ where the density of states becomes close to 1 sequence, or when $q^L \Omega(E) \sim 1$. Substituting the mean evolved sequence energy $E = \langle E \rangle  - \chi^2$ and taking the log, this is approximately when $\chi^2 \approx 2 L \log q$, or $\langle \chi_i^2 \rangle \approx 2 \log q$.

\section{\label{regstrength} Solving for $\gamma^{ij}$ Numerically}

As described in the main text we choose the regularization strengths $\gamma^{ij}$ such that the biased bivariate marginals $\tilde{f}^{ij}_{\alpha_\beta}$ are likely to have generated the observed bivariate marginals $\hat{f}^{ij}_{\alpha_\beta}$ by chance due to finite sampling. For each position-pair $i,j$ we solve for the maximum value of $\gamma^{ij}$ which satisfies the inequality
\begin{equation}
E[\text{KL}(F_{\alpha\beta}, \tilde{f}_{\alpha\beta})] \ge \text{KL}(\hat{f}_{\alpha\beta}, \tilde{f}_{\alpha\beta})
\label{eq:gamma_equality}
\end{equation}
where $F_{\alpha\beta}$ are sample marginals drawn from a multinomial distribution around $\tilde{f}_{\alpha\beta}$ with sample size $N$. This equality can be solved by various numerical strategies, but many of these are computationally costly. The main difficulty is in evaluating the expectation value. Here we describe a fast and accurate approximation. 

Consider a pair $i,j$, dropping the $ij$ indexes here. We want to evaluate $E[\text{KL}(\tilde{F}_{\alpha_\beta}, \tilde{f}_{\alpha_\beta})] = \sum_{\alpha\beta} E[x_{\alpha\beta} \log x_{\alpha\beta}] - \sum_{\alpha\beta} \tilde{f}_{\alpha\beta} \log \tilde{f}_{\alpha_\beta}$, where the expectation value averages over a multinomial distribution for a sample of size $N$, and $x_{\alpha\beta}$ is the sampled marginal from $\tilde{f}$ (with $x_{\alpha\beta} = n/N$ for integer sample $n$), and we have used the multinomial expectation $E[x_{\alpha\beta}] = \tilde{f}_{\alpha\beta}$. The first term is the expectation of an entropy, which is simplified as
\begin{widetext}
\begin{align}
E[x_{\alpha\beta} \log x_{\alpha\beta}] &= \sum_{n=0}^N \frac{N!}{(N-n)! n!} \tilde{f}_{\alpha\beta}^n (1-\tilde{f}_{\alpha\beta})^{N-n} \frac{n}{N} \log \frac{n}{N} \\
&= \tilde{f}_{\alpha\beta} \sum_{n=1}^N \frac{(N-1)!}{(N-n)! (n-1)!} \tilde{f}_{\alpha\beta}^{n-1} (1-\tilde{f}_{\alpha\beta})^{N-n} \log \frac{n}{N} \\
&= \tilde{f}_{\alpha\beta} \sum_{m=0}^N \frac{M!}{(M-m)! m!} \tilde{f}_{\alpha\beta}^m (1-\tilde{f}_{\alpha\beta})^{M-m} \log \frac{m+1}{M+1} \\
&= \tilde{f}_{\alpha\beta} E_M[\log \frac{m+1}{M+1}]
\end{align}
\end{widetext}
where $M=N-1$, $m = n-1$, and the last expectation value is over a binomial distribution with $M$ samples. Next we use a Taylor approximation $E[\log(x)] \approx \log(E[x]) - \frac{V[x]}{2 E[x]^2}$, and find that
\begin{widetext}
\begin{align}
E[\log \frac{m+1}{M+1}] &\approx \log (\tilde{f}_{\alpha\beta} + \frac{1-  \tilde{f}_{\alpha\beta}}{N}) - \frac{(N-1) \tilde{f}_{\alpha\beta} (1-\tilde{f}_{\alpha\beta} )}{2((N-1)  \tilde{f}_{\alpha\beta} + 1)^2} \\
E[\text{KL}(F_{\alpha_\beta}, \tilde{f}_{\alpha\beta})] &\approx \sum_{\alpha\beta} \tilde{f}_{\alpha\beta} \left( \log (\tilde{f}_{\alpha\beta} + \frac{1-  \tilde{f}_{\alpha\beta}}{N} )-  \frac{N-1}{2 N^2} \frac{ \tilde{f}_{\alpha\beta} (1-\tilde{f}_{\alpha\beta} )}{(\tilde{f}_{\alpha\beta} + \frac{1-  \tilde{f}_{\alpha\beta}}{N})^2} \right).
\end{align}
\end{widetext}

This gives us a way to quickly evaluate equation \ref{eq:gamma_equality} for any choice of $\gamma$, and we can then minimize the left-hand-side numerically by any standard method. We find this approximation is very good in practice.

\section{\label{convergence} Convergence}

\begin{figure*}
\centering
\includegraphics[width=\columnwidth]{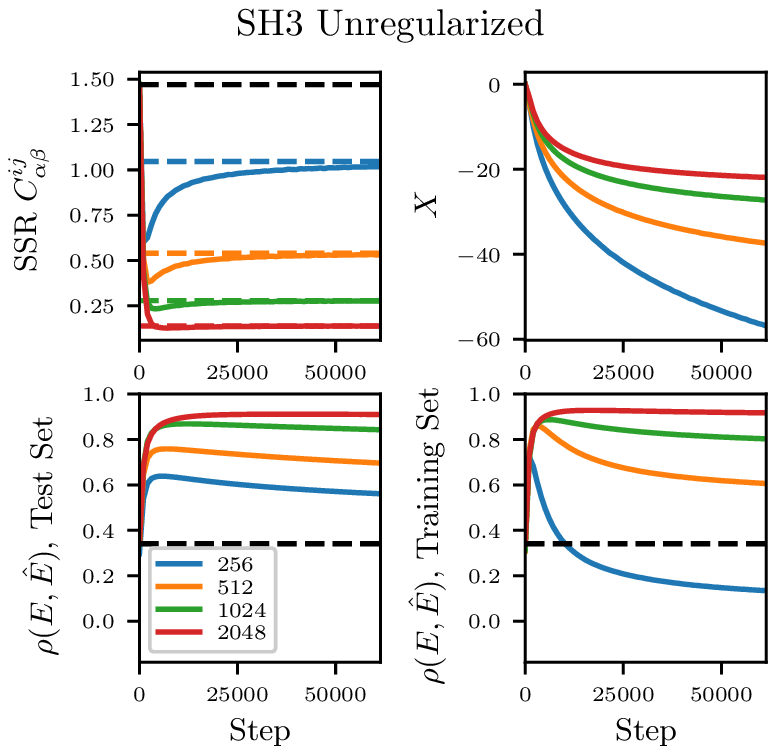}
\includegraphics[width=\columnwidth]{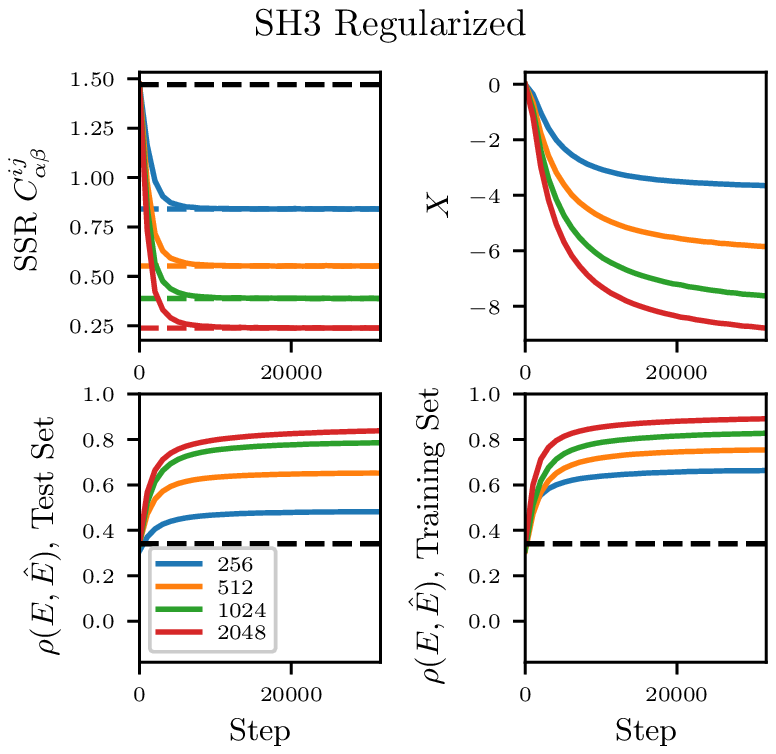}
\includegraphics[width=\columnwidth]{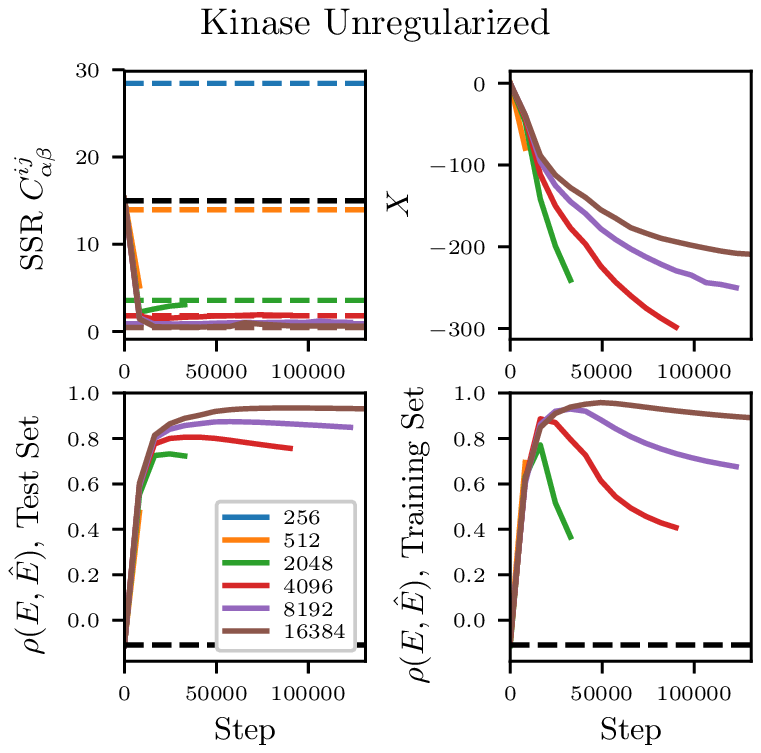}
\includegraphics[width=\columnwidth]{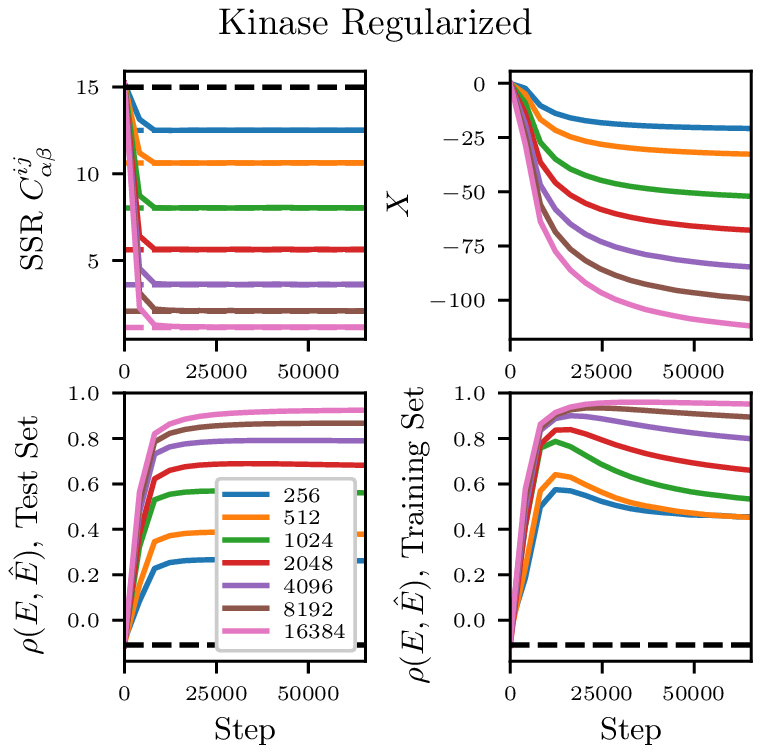}
\caption{Inverse Ising inference equilibration for different \textit{in-silico} MSA depths $N$, for different datasets. In all plots the x-axis shows number of coupling-updates during inference. For each of the models there are four subplots. Upper left subplots: Error in the model's correlation statistics relative to the reference model, measured as the sum of squared residuals (SSR) of the correlation coefficients $C^{ij}_{\alpha \beta}$. Dotted lines are the SSR of the training MSA relative to the reference, plus the independent model in black. Upper right subplots: Total correlation energy $X$ as a function of step. The independent model has $X=0$ by definition.  Lower left supblots: $\rho(E, \hat{E})$ for a test MSA of 4096 sequences drawn from the reference model. The black dotted line is the correlation of the independent model. Lower right subplots: $\rho(E, \hat{E})$ when scoring the training MSA sequences.}
\label{fig:equil}
\end{figure*}

In this appendix we show plots illustrating the convergence of the inverse Ising inference for the kinase and SH3 datasets, in figure \ref{fig:equil}.

\end{document}